# What do the US West Coast Public Libraries Post on Twitter?

**Amir Karami**
*University of South Carolina, USA. karami@sc.edu*

**Matthew Collins**
*University of South Carolina, USA. mcollinsga@hotmail.com*

## ABSTRACT

Twitter has provided a great opportunity for public libraries to disseminate information for a variety of purposes. Twitter data have been applied in different domains such as health, politics, and history. There are thousands of public libraries in the US, but no study has yet investigated the content of their social media posts like tweets to find their interests. Moreover, traditional content analysis of Twitter content is not an efficient task for exploring thousands of tweets. Therefore, there is a need for automatic methods to overcome the limitations of manual methods. This paper proposes a computational approach to collecting and analyzing using Twitter Application Programming Interfaces (API) and investigates more than 138,000 tweets from 48 US west coast libraries using topic modeling. We found 20 topics and assigned them to five categories including public relations, book, event, training, and social good. Our results show that the US west coast libraries are more interested in using Twitter for public relations and book-related events. This research has both practical and theoretical applications for libraries as well as other organizations to explore social media actives of their customer and themselves.

## INTRODUCTION

In the 21st century, libraries of all types are responding to the changing social, economic, and political impacts of living in a digital society. There are more than 9,000 public libraries[1] in the US that provide vital community services, such as early childhood literacy, computer training, and workforce development; they also invite community conversations and actions that further understanding and address local needs. In addition, public libraries act as a safe place for all patrons, reflecting and serving the diversity of their communities in their collections, programs, and services (ALA, 2014). In many communities, libraries advise patrons on issues; ranging from navigating the health system to helping those with housing needs. This “go-to” role has influenced library programming and events, with libraries providing advice and connections to health, housing, literacy, and other areas (Cabello & Butler, 2017). According to a 2015 Pew survey, almost two-thirds of Americans said that closing their local library would have a major impact on their community. Over 90 percent of adults think of public libraries as “welcoming and friendly places,” and about half have visited or otherwise used a public library in the last 12 months (Zickuhr, Rainie, Purcell, & Duggan,2013).

Recent research indicates that more people are using social media applications. According to Statista Dossier (2014), the number of social network users will increase from 0.97 billion to 2.44 billion users in 2018, predicting an increase around 300% in 8 years. Considering its rapid development, social media may become the most important media channel for organizations to reach their clients in the near future (He, Zha, & Li, 2013; Moro, Rita, & Vala, 2016). Since its launch in July 2006, Twitter has quickly become one of the most popular social networking platforms for users to update their followers, and to provide convenient and effective information dissemination (Del Bosque, Leif, & Skarl, 2012; Shulman, Yep, & Tomé, 2015). Organizations have realized the potential of using internet-based social networks to influence customers, incorporating social media marketing communication in their strategies for leveraging their businesses. Measuring the impact of messaging is an important process for developing an effective social media strategy (Moro et al., 2016).

Beyond continued access to traditional collections and services, public libraries engage their patrons and communities through several new participation platforms, such as social media and maker spaces, which enable and mediate new forms of online user participation, engagement, access, and interactivity (Cavanagh, 2016). Social media applications have been increasingly adopted by libraries to market both resources and services as well as to enhance the relationship with their patrons. It enables "information and knowledge sharing, service enhancement and promotion, interaction with student library users, at minimal costs" (Collins & Quan-Haase, 2014; Chu & Due, 2013). Social media functions as a useful tool for libraries to reach out to current and prospective patrons outside of the physical library setting. Using social media can enable libraries to distribute information through a medium with which many patrons are already familiar. Libraries must ensure the forms of social media they employ and the subsequent content they produce are relevant to the interests and information-based needs of their audience (Collins & Quan-Haase, 2014).

[1] http://www.ala.org/tools/libfactsheets/alalibraryfactsheet01

Research has shown that when a library has an engaged and active Twitter following, it is able to spread information more easily (Yep, Brown, Fagliarone, and Shulman, 2017). Twitter shows both sides of user evolution; as users offer their opinions and needs, libraries have real-time feedback they can utilize to adjust their services. Simultaneously, Twitter gives libraries unprecedented access to their users. Social media in general allows libraries to engage users who may otherwise never think of using the library's services. In addition to proactive community outreach programs, Twitter enables libraries to advertise their services and programs to many more users than traditional media. Mining these data provides an instantaneous snapshot of the content of public library tweets (Chew and Eysenbach, 2010). Social media sheds light on how patrons use library services, as well as assists academic libraries in advancing research (Yep, Brown, Fagliarone, and Shulman, 2017).

Apart from facilitating discussion between libraries, social media analysis (SMA) can also help libraries and other non-profit institutions to measure the satisfaction level of their users. When using SMA to their advantage, organizations are able to tailor their products to the needs of their clients and patrons. This is especially affordable for small and mid-size libraries that do not have the financial resources to afford traditional marketing campaigns. In addition to lower costs, libraries improve on serving their patrons when they utilize social media.

Twitter data analysis provides a convenient means of finding data for an organic user study. Twitter data have been applied in different domains such as health (Shaw, & Karami, 2017), business (Bollen, Mao, & Zeng, 2011), and politics (Karami, Bennett, & He, 2018; Najafabadi & Domanski, 2018; Najafabadi, 2017). For scholars who are looking for real-time conversational or text data, Twitter has provided a wealth of data for behavioral analysis (Karami, Bennett, & He, 2018), sentiment analysis (Shaw & Karami, 2017), trend analysis, information dissemination, and health surveillance (Karami, Dahl, Turner-McGrievy, Kharrazi, & Shaw, 2018; Webb, Karami, & Kitzie, 2018). Analyzing this vast amount of data can reveal knowledge and interesting patterns that can lead to new discoveries (He, Zha, & Li, 2013).

Although previous studies have provided valuable insights to social media analysis, other studies have yet to investigate the content of tweets posted by libraries to disclose their interests. In this study, we mine and analyze public library-related topics extracted from the Twitter accounts of 48 public libraries in the western US in order to examine how public libraries use social media to describe their services and interact with patrons. In the present study, we propose a computational approach to collect and analyze more than thousands of tweets to discover meaningful themes and semantic patterns in the data. This paper specifically addresses this question: *What do the US west coast public libraries post on Twitter?*

## RELATED WORKS

Several previous studies have analyzed the content of social media posts. Twitter is one of the most popular social media applications used by studies from diverse fields, such as mental health (Jamison-Powell et al., 2012), libraries (Collins & Karami, 2018; Stvilia & Gibradze, 2017), and journalism and mass communication (Guo, Vargo, Pan, Ding, & Ishwar, 2016; Moro et al., 2016).

In an effort to help companies understand how to perform a social media competitive analysis and transform social media data into knowledge for decision makers and e-marketers, one study applied text mining to unstructured text content on Facebook and Twitter sites of the three largest pizza chains: Pizza Hut, Domino's Pizza and Papa John's Pizza. The authors manually collected data and utilized SPSS Clementine and Nvivo 9 for data analysis. Results revealed that the pizza chains actively used social media and have committed substantial resources for their social media efforts. The study also demonstrated that the three largest pizza chains have made significant social media efforts to increase interaction with customers and build brands in the online communities (He, Zha, & Li, 2013).

From another study that explored the role of social media in the discussion of mental health issues, and with particular reference to insomnia and sleep disorders, one research collected tweets using a tweet archiving application, but analyzed tweet content through a mixed methods approach. First, the Linguistic Enquiry and Word Count (LIWC) software performed an automated sentiment and linguistic content analysis, then the authors conducted a hand-coded, inductive thematic analysis. From the analysis six themes were identified. The study many several observations based on their content analysis, but unfortunately did not compare the automated and manual methods, nor state a methodological preference (Jamison-Powell, Linehan, Daley, Garbett, & Lawson, 2012). LIWC is a tool for capturing thoughts, feelings, personality, and motivations, to the dataset (Karami & Zhou, 2015; Karami & Zhou, 2014b).

Many studies have examined the content of library tweets, as well as the effectiveness of Twitter usage by libraries (Al-Daihani & AlAwadhi, 2015; Yep, Brown, Fagliarone, & Shulman, 2017). Aharony (2010) explored the use of Twitter in public and academic libraries to understand microblogging patterns. The study focused its research on three areas: (1) Do public libraries produce more or less tweets than academic libraries, (2) Is there a linguistic difference between tweets produced by public libraries and those produced by academic libraries, (3) Is the content of the tweets produced by public libraries different from that of tweets produced by academic libraries? Once data was collected manually, it was first, manually analyzed through a statistical descriptive analysis that described the basic features of the data in a study, and then secondly, manually examined with inferential content analysis which divided the tweets by topic categories and sub-categories. Results showed that the total number of tweets in public libraries was larger than the total number of tweets in academic libraries, most of the tweets posted by academic libraries used formal language and public libraries used informal language, and both kinds of libraries use Twitter to broadcast and share information about their activities, opinions, status, and professional interests.

## METHODOLOGY AND RESULTS

This paper proposes a computational framework for mining the tweets of libraries to disclose their social media content in Twitter. This framework has four steps including Twitter account detection, tweets collection, topic discovery, and topic analysis (Figure 1).

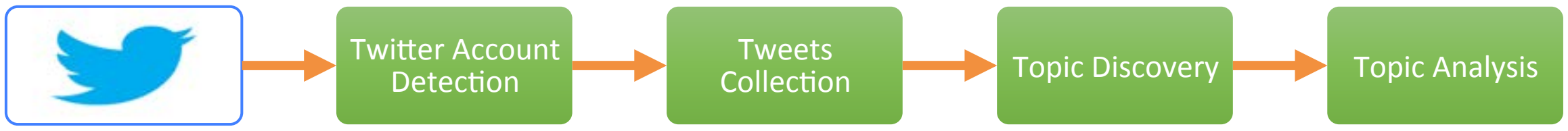

Figure 1: Research Framework

### *Twitter Account Detection*

The first step in this study is to detect the Twitter accounts of libraries in the western US west coast. We searched in Google to find the names of libraries in the five west coast states, including Alaska (AK), California (CA), Hawaii (HI), Oregon (OR), and Washington (WA). Then, we went to the websites of the libraries to find their Twitter account. In order to focus on active Twitter accounts, we selected accounts having at least 100 tweets such as the examples in Figure 1. Finally, we found 48 active Twitter accounts across five states, including 5 active Twitter accounts in Alaska; 22 in California; 1 in Hawaii; 9 in Oregon; and 11 in Washington (Table 1).

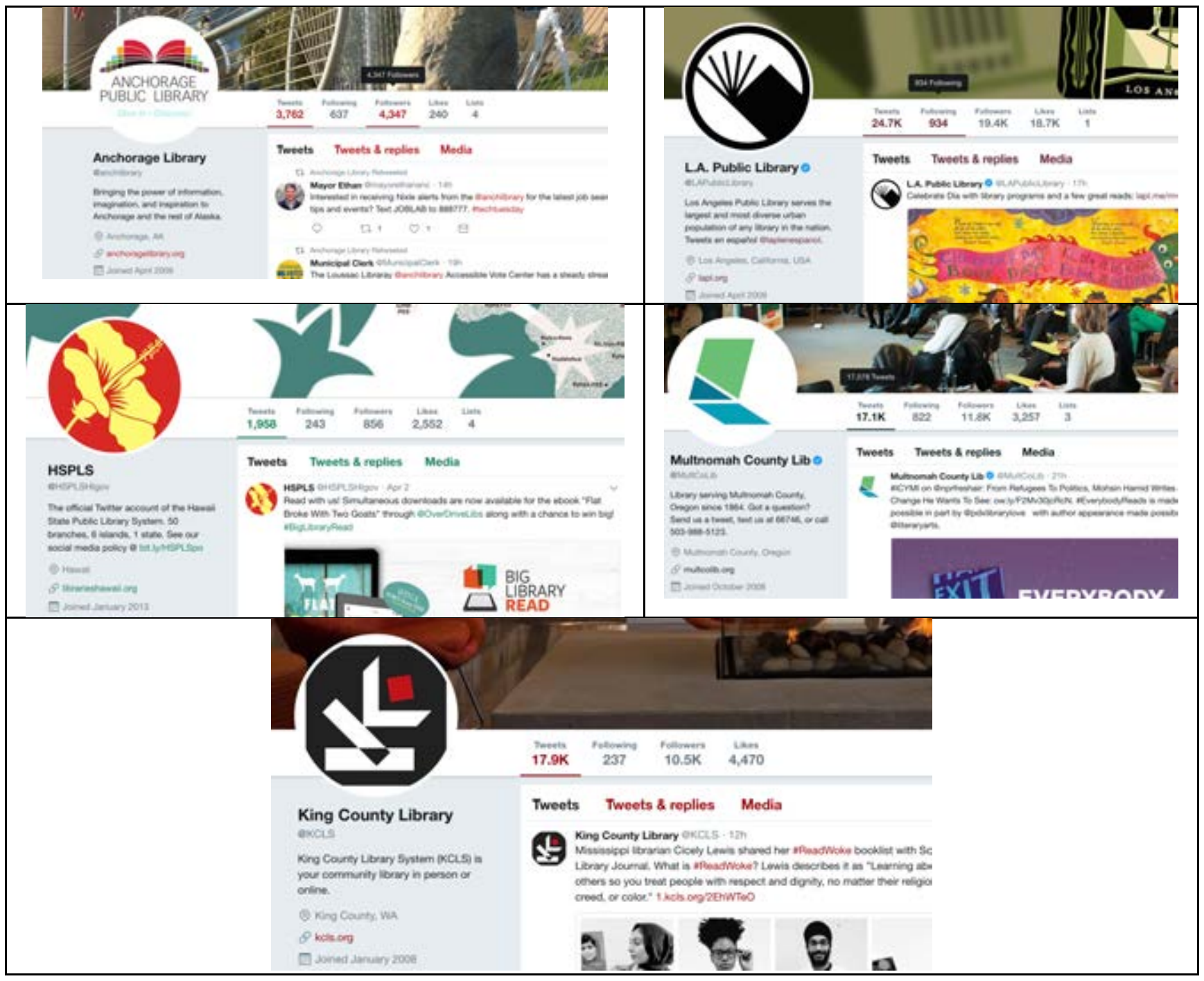

**Figure 1: Twitter Accounts of 5 West Coast Libraries**

| State | Library Name | State | Library Name |
|---|---|---|---|
| **Alaska** | Anchorage Public Library | **Hawaii** | Hawaii State Public Library |
| | Ketchikan Public Library | **Oregon** | Multnomah County Library |
| | Kodiak Public Library | | Beaverton City Library |
| | Kuskokwim Consortium Library | | Cedar Mill Community Library |
| | Homer Public Library | | Jackson County Library Services |
| **California** | Los Angeles Public Library | | Salem Public Library |
| | San Diego Public Library | | Eugene Public Library |
| | San Jose Public Library | | Hillsboro Public Library |
| | Fresno County Public Library | | Corvallis-Benton County Public Library |
| | Orange County Public Libraries | | Washington County Cooperative Library Services |
| | San Francisco Public Library | **Washington** | King County Library System |
| | Marin County Free Library | | Sno-Isle Libraries |
| | Solano County Library | | Seattle Public Library |
| | Pasadena Public Library | | Pierce County Library |
| | Santa Cruz Public Libraries | | Edmonds Library |
| | San Mateo County Libraries | | Timberland Regional Library |
| | Santa Monica Public Library | | North Central Regional Library |
| | Monterey Free County Libraries | | Tacoma Public Library |
| | Stanislaus County Library | | Spokane County Library District |
| | Long Beach Public Library | | Bellingham Public Library |
| | Santa Clara County Library District | | Whatcom County Library System |
| | Sacramento Public Library | | |
| | Alameda County Library | | |
| | Thousand Oaks Library | | |
| | Riverside County Library System | | |
| | Fortuna Library | | |
| | San Diego County Library | | |

**Table 1: The List of West Coast Libraries**

### *Tweets Collection*

To collect the tweets of each of the 48 libraries, we used the Twitter Application Programming Interfaces (API) in R platform[2]. Due to the API limit, we could only extract up to 3,200 tweets per account. After removing the duplicate tweets, we collected 138,056 tweets with 2,876.17 tweets per account (Table 2). Although California had the highest number of libraries in our dataset, based on average tweets/library score, it seemed that the Alaska and Oregon libraries were more active on Twitter. It is worth mentioning that this data will be publicly available in the first author's webpage[3].

---

[2] https://cran.r-project.org/web/packages/twitteR/twitteR.pdf
[3] xxx

| State | Number of Tweets | Number of Libraries | Average Tweets/Library |
|---|---|---|---|
| **AK** | 26,739 | 5 | 5,347.8 |
| **CA** | 65,153 | 22 | 2,961.5 |
| **HI** | 1,739 | 1 | 1,739 |
| **OR** | 29,166 | 9 | 3,240.7 |
| **WA** | 24,259 | 11 | 2,205.36 |
| **Total** | 138,056 | 48 | 2,876.17 |

**Table 2: Statistics of Tweets**

### *Topic Discovery*

Traditional concept analysis approach is not an efficient strategy to understand thousands of tweets (Karami, 2017). Therefore, there is a need to use computational methods to explore a huge number of social media posts (Karami, Gangopadhyay, Zhou, & Kharrazi, 2015a). We utilized topic modeling to discover topics in the collected tweets. Among different topic models, Latent Dirichlet Allocation (LDA) is an efficient machine learning method for text mining. LDA assumes that there are topics in a document and words can be assigned to the topics with different weights (Blei, Ng, & Jordan, 2003; Karami, 2015). This model categorizes the words that are semantically related in a topic and represented thematically. For example, this model assigns "life," "evolve," and "organism" to a topic with Evolution theme (Figure 2).

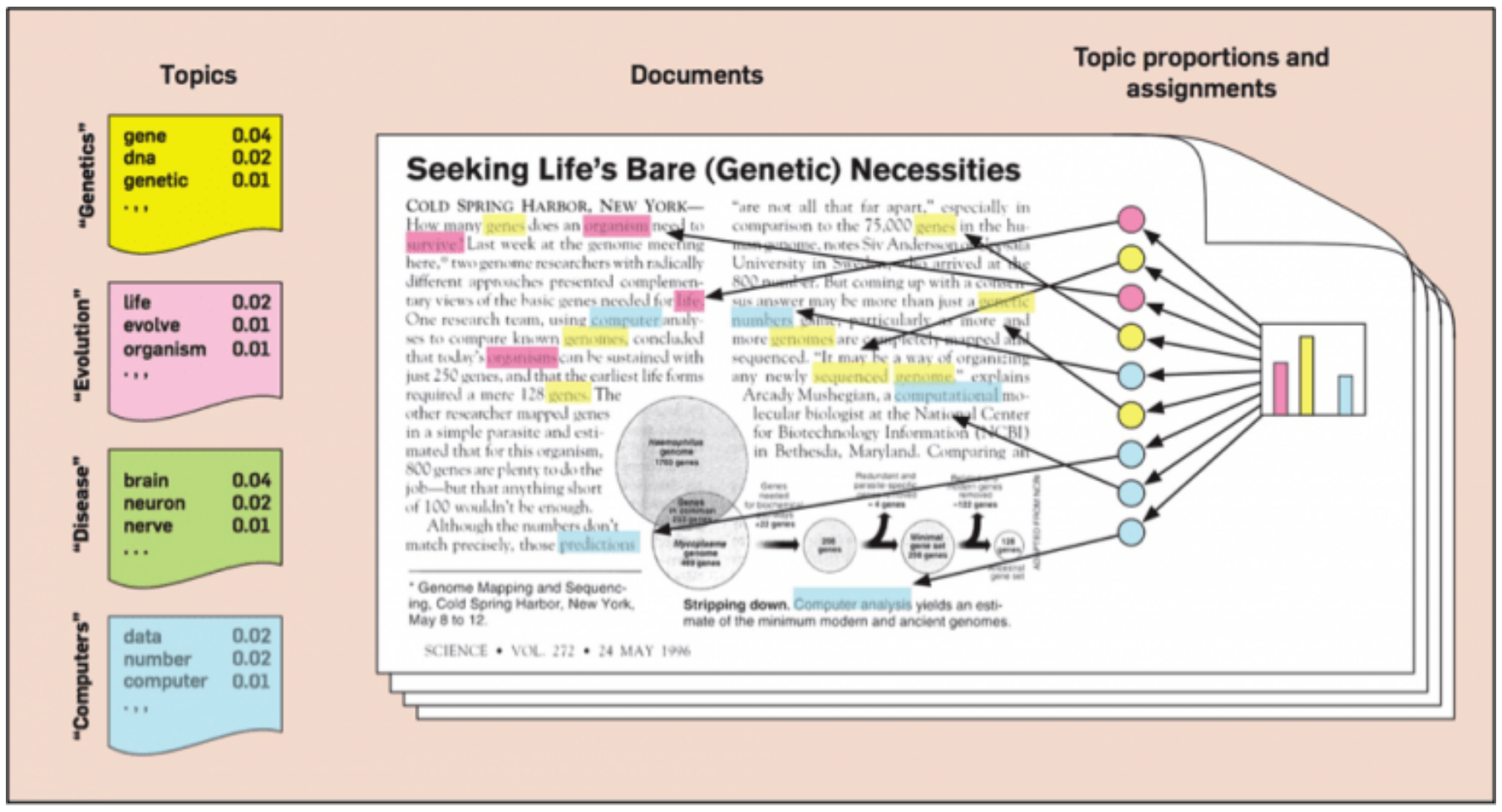

**Figure 2: The Intuition behind LDA (Blei, 2012)**

LDA has been used for different applications, such as medical research (Ghassemi, Naumann, Doshi-Velez, Brimmer, Joshi, Rumshisky, & Szolovits, 2014; Karami, Gangopadhyay, Zhou, & Kharrazi, 2018; Karami, Gangopadhyay, Zhou, & Karrazi, 2015b; Karami & Gangopadhyay, 2014), spam detection (Karami & Zhou, 2014a), and business (Karami & Pendergraft, 2018). While this text mining method has been applied on Twitter data, such as analyzing health tweets (Ghosh & Guha, 2013) and politics (Karami, Bennett, & He 2018), this model has not been considered for analyzing library tweets to disclose topics of interest in their social media posts. We applied Mallet Java-based implementation of LDA (McCallum, 2002). Mallet removed the stopwords such as "the" and "a" that do not have semantic value for our analysis. After removing the stopwords, such as "the" and "a" that don't have semantic value for our analysis, we found 20 topics (Table 3).

For *138,056* documents (tweets) and *20* topics, LDA also measured the probability for each of the topics in each of the documents or $P(T_k|D_j)$. We used $P(T_k|D_j)$ to find the weight of each of topics, $WT(T_k)$. For an effective comparison, each of WTs was normalized by the sum of the weight scores of all topics:

$$\text{N_WT}(T_k) = \frac{\sum_{j=1}^{138,056} \text{P}(T_k|D_j)}{\sum_{k=1}^{20} \sum_{j=1}^{138,056} \text{P}(T_k|D_j)}$$

If $N_WT(T_x) > N_WT(T_y)$, it means that libraries posted more tweets about topic *x* than topic *y*. The first column of Table 3 shows the ranking of topics based on the weight of topics in Figure 3.

### *Topic Analysis*

In this step, we labeled the associated words based on overall theme in each of the topics (Table 3). For example, “facebook,” “posted,” “photos,” “online,” and “exhibit” represents a topic related to “social media activities”. This topic indicates the activities of the libraries in social media websites, such as Facebook. Table 3 also shows our interpretation and description for each of the topics in the second column. Using $WT(T_k)$ helped us to measure the weight of the topics and determine the rank for the libraries. While book recommendations, book reviews, and library hours are the three most frequent topics, health education, public service, and public talks are the least frequent according to our topics’ ranking (Figure 3).

| Rank | Associated Words for Topic<br>Topic Description | Label |
|---|---|---|
| 1 | Associated words: books, read, book, list, favorite<br>Describing books recommended by library staff for patrons. | Book Recommendations |
| 2 | Associated words: new, book, author, life, story<br>Reviewing of the quality, author, plot, or subject of a book. | Book Reviews |
| 3 | Associated words: closed, open, library, hours, locations, branches<br>Describing a library’s hours of operation. | Library Hours |
| 4 | Associated words: free, learn, workshop, class, job, register<br>Describing a training session or workshop at the library. | Public Trainings |
| 5 | Associated words: card, ebooks, collection, catalog, access, audiobooks<br>Promoting the library’s services, resources, and collection. | Promotions of Library Services |
| 6 | Associated words: great, awesome, staff, people, good, folks<br>Describing gratitude or appreciation for staff, volunteers, or patrons. | Appreciation for Events |
| 7 | Associated words: county, community, public, support, libraries, board<br>Requesting community and patron support for the library. | Request for Support |
| 8 | Associated words: free, movie, night, family, fun, games<br>Describing a fun or enjoyable event at the library. | Fun Events |
| 9 | Associated words: summer, reading, program, kids, read<br>Describing the upcoming summer reading program at a library. | Summer Reading Program |
| 10 | Associated words: book, author, join, local, event<br>Describing a book-related event scheduled to occur at a library. | Book Events |
| 11 | Associated words: book, friends, big, sale, today<br>Describing a book sale occurring at the library. | Book Sale |
| 12 | Associated words: teen, event, club, tonight, scavenger, hunt<br>Describing a library event intended for teenagers. | Teen Events |
| 13 | Associated words: happy, celebrate, birthday, national, day, anniversary<br>Describing an upcoming celebration to promote a program, issue, or campaign. | Theme Days and Months |
| 14 | Associated words: event, public, today, city, park<br>Describing an event scheduled to occur nearby. | Local Events |
| 15 | Associated words: music, free, concert, tonight, rock, jazz<br>Describing an upcoming musical performance or concert. | Music Events |
| 16 | Associated words: story, time, kids, ages, children’s, stories<br>Describing a library event planned for children. | Children’s Events |
| 17 | Associated words: facebook, posted, photos, online, exhibit<br>Describing information posted to a library’s social media account. | Social Media Activities |
| 18 | Associated words: healthy, life, food, learn, health, garden<br>Describing issues and resources for healthy living. | Health Education |
| 19 | Associated words: business, tax, learn, resources, free, great<br>Describing resources and services provided by the library. | Public Services |
| 20 | Associated words: learn, today, talk, join, news<br>Describing an upcoming talk or lecture at the library. | Public Talks |

**Table 3: Library Topic Descriptions**

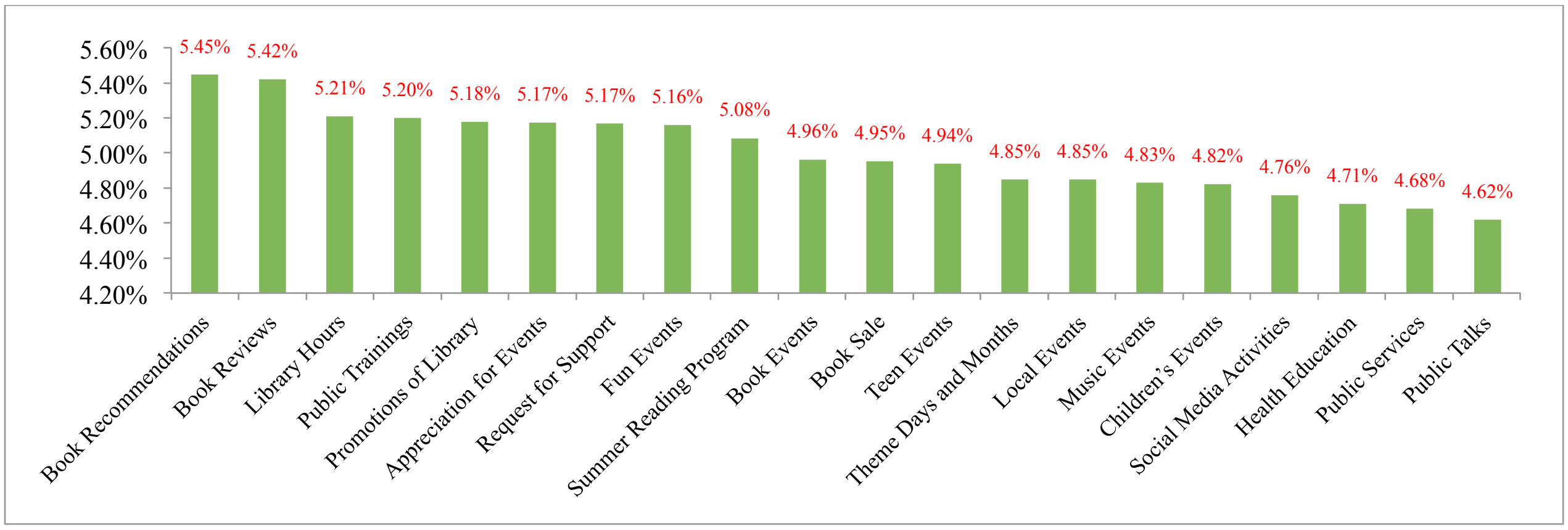

**Figure 3: Weight of Topics**

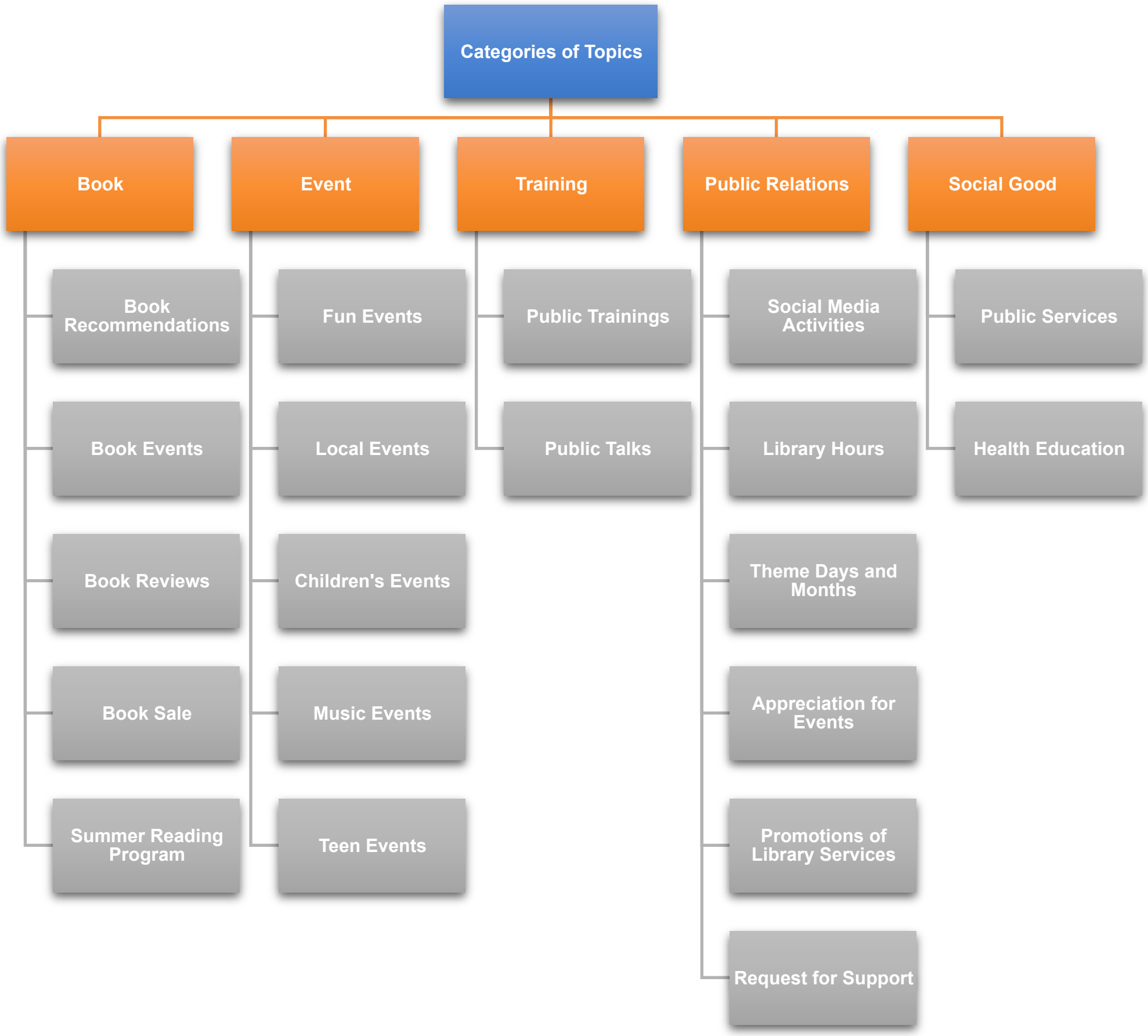

**Figure 4: Categories of Topics**

We also observed that the 20 topics could be assigned to five categories representing multiple topics including book, event, training, public relations, and social good (Figure 4):

1. The **Book** category shows the topics that book-related issues including book recommendations, book events, book reviews, book sale, and summer reading program.
2. The **Event** category describes library and local events including fun events, local events, children's events, music events, and teen events. The book events was assigned to the book category.
3. The **Training** category represents two topics including public trainings and public talks. This category is about library and non-library trainings, workshops, lectures, or talks occurring including public trainings and public talks.
4. The **Public Relations** category describes topics regarding the library public communication. The six related topics

are social media activities, library hours, theme days and months, appreciation for events, promotions of library services, and request for support.

5. The **Social Good** category covers the topics related to issues benefiting the entire community such as tax filing. The topics are related to this category: library services and health education.

We also used $N_WT(T_k)$ to find the weight of categories for each state. This means that we combined the weight of topics for each category and then found the weight of the category and normalized it. For example, we measured the total weight of public training (PTr) and public talks (PTa) for the libraries in Oregon. Then calculated the sum of PTr and PTa (PTr + PTa) as the weight of training category for Oregon. Figure 5 shows the weight of each of the five categories for each state along with the total weight of categories using a west cost (WC) label. It seems that around 55% of tweets were about public relations and book categories. We also found that:

- *Public relations*, *event*, and *book* were the top three categories in four states; in Alaska, the event category had a higher weight than the book category. These three categories were discussed in around 80% of the tweets.
- There is no specific pattern for training and social good categories. While Alaska, Hawaii, and Oregon libraries preferred training more than social good in their Twitter posts, libraries in California and Washington were vice versa. The total weight of these two categories was around 20%.

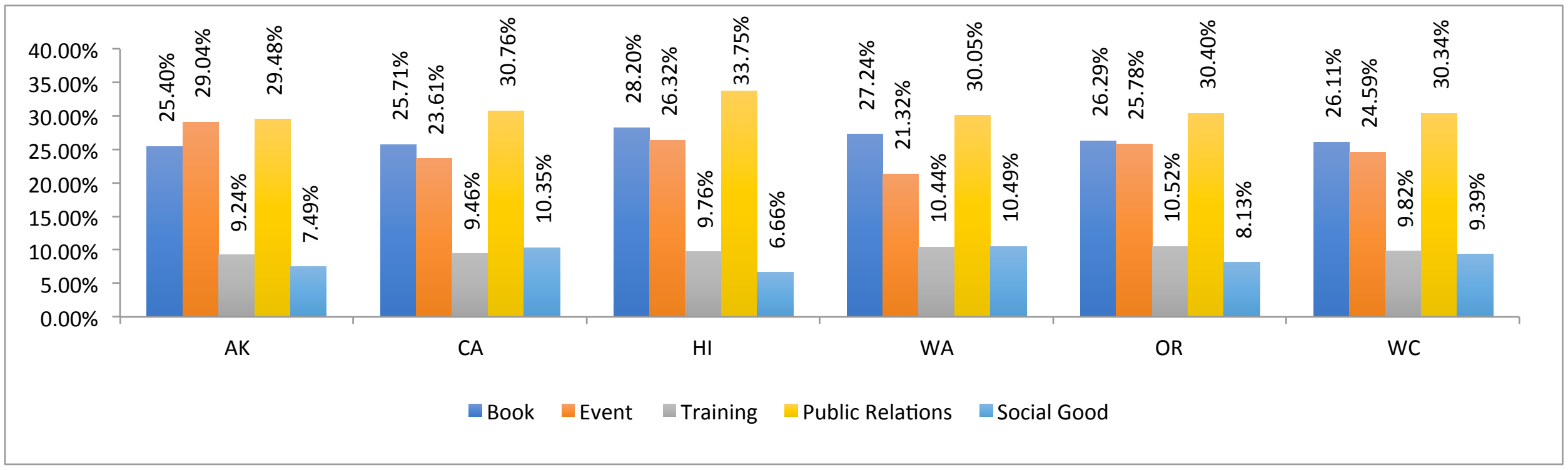

**Figure 5: Weight of Categories per State**

## CONCLUSION

Twitter is a popular social media for different organizations to disseminate information for their followers. There is a wide range of studies to investigate what people and organizations post on Twitter; however, there is no study to detect the interests of public libraries in Twitter. This study proposed a computational framework to collect and explore thousands of tweets to disclose hidden semantic structure in them. We used LDA to find 20 topics and grouped them into five categories. LDA also helped us to find the weight of topics across the five states in our corpus.

The methodology and results in this paper have practical applications for library studies. Libraries as well as other organizations can reap benefit from this study by computational analysis of their social media content. Our future research plans will consider user location and time of tweet as added layers for investigating Twitter accounts of other US and non-US libraries.

## ACKNOWLEDGMENTS

This research is supported in part by the Provost Office Social Science Research Grant and the first author's startup research funding provided by the School of Library and Information Science at the University of South Carolina. All opinions, findings, conclusions and recommendations in this paper are those of the authors and do not necessarily reflect the views of the funding agency.

## REFERENCES

Al-Daihani, S. M., & AlAwadhi, S. A. (2015). Exploring academic libraries ’ use of Twitter : a content analysis. *The Electronic Library*, *33*(6), 1002–1015. https://doi.org/10.1108/EL-05-2014-0084

American Libraries Association (ALA). (2014). *State of America’s Libraries Report 2014*. Retrieved from

http://www.ala.org/news/state-americas-libraries-report-2014

Blei, D. M., Ng, A. Y., & Jordan, M. I. (2003). Latent Dirichlet Allocation. *Journal of Machine Learning Research*, *3*, 993–1022.

Blei, D. M. (2012). Probabilistic topic models. Communications of the ACM, 55(4), 77-84.

Bollen, J., Mao, H., & Zeng, X. (2011). Twitter mood predicts the stock market. *Journal of Computational Science*, 2(1), 1-8.

Cabello, M., & Butler, S.M. (2017, March 30). How public libraries help build healthy communities? *The Brookings Institution*. Retrieved from https://www.brookings.edu/blog/up-front/2017/03/30/how-public-libraries-help-build-healthy-communities/

Cavanagh, M. F. (2016). Micro-blogging practices in Canadian public libraries : A national snapshot. *Journal of Librarianship and Infomation Science*, *48*(3), 247–259. https://doi.org/10.1177/0961000614566339

Chew, C., & Eysenbach, G. (2010). Pandemics in the Age of Twitter : Content Analysis of Tweets during the 2009 H1N1 Outbreak. *PLoS ONE*, 5(11), 1–13. https://doi.org/10.1371/journal.pone.0014118

Chu, S., & Du, H. (2013). Social networking tools for academic libraries. *Journal of Librarianship and Information Science*, 45(1), 64-75. https://doi.org/10.1177/0961000611434361

Collins, G., & Quan-Haase, A. (2014). Are social media ubiquitous in academic libraries? A longitudinal study of adoption and usage patterns. *Journal of Web Librarianship*, 8(1), 48–68. https://doi.org/10.1080/19322909.2014.873663

Collins M., & Karami A. (2018), Social Media Analysis for Organizations: US Northeastern Public and State Libraries Case Study, Proceedings of the Southern Association for Information Systems (SAIS), Atlanta, GA.

Del Bosque, D., Leif, S. A., & Skarl, S. (2012). Libraries atwitter: trends in academic library tweeting. *Reference Services Review*, 40(2), 199–213. https://doi.org/10.1108/00907321211228246

Dossier, Statista (2014). Social media & user-generated content—Number of global social network users 2010-2018-Statista Dossier 2014. Retrieved September 10, 2015, from http://www.statista.com/statistics/278414/number-of-worldwide-socialnetwork-users/

Gao, W., & Wallace, L. (2017). Data Mining , Visualizing , and Analyzing Faculty Thematic Relationships for Research Support and Collection Analysis. In *ACRL 2017* (pp. 171–178). Baltimore.

Ghosh, D., & Guha, R. (2013). What are we 'tweeting'about obesity? Mapping tweets with topic modeling and Geographic Information System. *Cartography and Geographic Information Science*, 40(2), 90-102.

Guo, L., Vargo, C. J., Pan, Z., Ding, W., & Ishwar, P. (2016). Big Social Data Analytics in Journalism and Mass Communication : Comparing Dictionary-Based Text Analysis and Unsupervised Topic Modeling. *Journalism & Mass Communication Quarterly*, 93(2), 332–359. https://doi.org/10.1177/1077699016639231

He, W., Zha, S., & Li, L. (2013). Social media competitive analysis and text mining : A case study in the pizza industry. *International Journal of Information Management*, 33, 464–472. https://doi.org/10.1016/j.ijinfomgt.2013.01.001

Jamison-Powell, S., Linehan, C., Daley, L., Garbett, A., & Lawson, S. (2012). "I can't get no sleep": Discussing #insomnia on Twitter. In *Proceedings of the SIGCHI Conference on Human Factors in Computing Systems* (pp. 1501–1510). Retrieved from http://andygarbett.co.uk/wp-content/uploads/2016/05/Jamison-Powell-et-al-I-Cant-Get-No-Sleep.pdf

Karami, A., Bennett, L. S., & He, X. (2018). Mining Public Opinion about Economic Issues: Twitter and the US Presidential Election. International Journal of Strategic Decision Sciences (IJSDS), 9(1), 18-28.

Karami, A., Dahl, A. A., Turner-McGrievy, G., Kharrazi, H., & Shaw, G. (2018). Characterizing diabetes, diet, exercise, and obesity comments on Twitter. International Journal of Information Management, 38, 1, 1-6.

Karami, A., Gangopadhyay, A., Zhou, B., & Kharrazi, H. (2018). Fuzzy approach topic discovery in health and medical corpora. International Journal of Fuzzy Systems, 20(4), 1334-1345.

Karami A., & Pendergraft N. M. (2018). Computational Analysis of Insurance Complaints: GEICO Case Study, International Conference on Social Computing, Behavioral-Cultural Modeling, & Prediction and Behavior Representation in Modeling and Simulation, Washington, DC.

Karami, A. (2017). Taming Wild High Dimensional Text Data with a Fuzzy Lash. In 2017 IEEE International Conference on Data Mining Workshops (ICDMW) (pp. 518-522). IEEE.

Karami, A. (2015). Fuzzy topic modeling for medical corpora. University of Maryland, Baltimore County.

Karami, A., Gangopadhyay, A., Zhou, B., & Kharrazi, H. (2015a). A fuzzy approach model for uncovering hidden latent semantic structure in medical text collections. iConference 2015 Proceedings.

Karami, A., Gangopadhyay, A., Zhou, B., & Karrazi, H. (2015b). Flatm: A fuzzy logic approach topic model for medical documents. In the Annual Conference of the North American Fuzzy Information Processing Society (NAFIPS) held jointly with the 5th World Conference on Soft Computing (WConSC), (pp. 1-6). IEEE.

Karami, A., & Zhou, B. (2015). Online review spam detection by new linguistic features. iConference 2015 Proceedings.

Karami, A., & Gangopadhyay, A. (2014). Fftm: A fuzzy feature transformation method for medical documents. Proceedings of BioNLP 2014, 128-133.

Karami, A., & Zhou, L. (2014a). Exploiting latent content based features for the detection of static sms spams. Proceedings of the American Society for Information Science and Technology, 51(1), 1-4.

Karami A., & Zhou L. (2014b). Improving Static SMS Spam Detection by Using New Content-based Features, Proceedings of the 20th Americas Conference on Information Systems (AMCIS), Savannah, GA.

Moro, S., Rita, P., & Vala, B. (2016). Predicting social media performance metrics and evaluation of the impact on brand building : A data mining approach. *Journal of Business Research*, *69*(9), 3341–3351. https://doi.org/10.1016/j.jbusres.2016.02.010

Najafabadi, M. M., & Domanski, R. J. (2018). Hacktivism and distributed hashtag spoiling on Twitter: Tales of the# IranTalks. First Monday, 23(4).

Najafabadi, M. M. (2017). A Research Agenda for Distributed Hashtag Spoiling: Tails of a Survived Trending Hashtag. In Proceedings of the 18th Annual International Conference on Digital Government Research (pp. 21-29). ACM.

Shaw, G., & Karami, A. (2017). Computational content analysis of negative tweets for obesity, diet, diabetes, and exercise. *Proceedings of the Association for Information Science and Technology*, 54(1), 357-365.

Shulman, J., Yep, J., & Tomé, D. (2015). Leveraging the Power of a Twitter Network for Library Promotion. *The Journal of Academic Librarianship*, 41(2), 178–185. https://doi.org/10.1016/j.acalib.2014.12.004

Stvilia, B., & Gibradze, L. (2014). Library & Information Science Research What do academic libraries tweet about , and what makes a library tweet useful ? *Library and Information Science Research*, *36*(3–4), 136–141. https://doi.org/10.1016/j.lisr.2014.07.001

Webb, F., Karami, A., & Kitzie, V. (2018). Characterizing Diseases and disorders in Gay Users' tweets. Proceedings of the Southern Association for Information Systems (SAIS), Atlanta, GA.

Yang, T.-I., Torget, A. J., & Mihalcea, R. (2011). Topic Modeling on Historical Newspapers. *Proceedings of the 5th ACL-HLT Workshop on Language Technology for Cultural Heritage, Social Sciences, and Humanities,* (June), 96–104.

Yep, J., Brown, M., Fagliarone, G., & Shulman, J. (2017). Influential Players in Twitter Networks of Libraries at Primarily Undergraduate Institutions. *The Journal of Academic Librarianship*, 43(3), 193–200. https://doi.org/10.1016/j.acalib.2017.03.005

Zickuhr, K., L. Rainie, K. Purcell, & M. Duggan. (2013). "How Americans value public libraries in their communities." *Pew Research Internet Project*. http://www.pewinternet.org/2014/03/13/library-engagement-typology/